\documentclass[aip, jmp, amsmath,amssymb, reprint]{revtex4-1}

\usepackage{graphicx}
\usepackage{dcolumn}
\usepackage{bm}

\begin{document}

	\title{A network of phase oscillators as a device for sequential pattern generation}
	
	\author{Pablo Kaluza \footnote{email: kaluza@fhi-berlin.mpg.de}}

	\affiliation{Abteilung Physikalische Chemie, Fritz-Haber-Institut der Max-Planck-Gesellschaft, Faradayweg 4-6, 14195 Berlin, Germany.}

	\affiliation{School of Engineering and Science, Jacobs University, P.O.Box 750561, 28725 Bremen, Germany.}

	\author{Hildegard Meyer-Ortmanns \footnote{email: h.ortmanns@jacobs-university.de}}
	\affiliation{School of Engineering and Science, Jacobs University, P.O.Box 750561, 28725 Bremen, Germany.}

	\date{\today}

	\begin{abstract}
	We design a system of phase oscillators that is able to produce temporally periodic sequences of patterns. Patterns are cluster partitions which encode information as phase differences between phase oscillators. The architecture of our system consists of a retrieval network with N globally coupled phase oscillators, a pacemaker that controls the sequence retrieval, and a set of patterns stored in the couplings between the pacemaker and the retrieval network. The system performs in analogy to a central pattern generator of neural networks and is very robust against perturbations in the retrieval process.
	\end{abstract}
	


	\maketitle 

	\begin{quotation}
	Sequential generation of patterns is an important issue in the context of dynamical systems that are designed for applications to artificial neural networks, gait models of animals, robot controlling and the like. In this paper, we construct a system of phase oscillators that is able to generate 
	periodic sequences of patterns via implemented couplings and an inherent clock. The system consists of a pacemaker providing the clock, a retrieval network providing the patterns, and a set of couplings between the pacemaker and the retrieval network, coding the patterns. The pacemaker activates the stored patterns one by one as a function of its own phase, and the retrieval network selects the cluster partition corresponding to the activated pattern. The process of selecting the cluster partition is realized by ensuring that the energy function of the retrieval network has only a single minimum while the pattern is activated. The gradient dynamics of this energy function then leads to a phase-locked motion of the retrieval network, such that the fixed phase relations between the oscillators represent the activated pattern. Thus the pattern amounts to a partition of oscillators into clusters characterized by a fixed phase relation (therefore called cluster partition). The system then performs similarly to a central pattern generator of biological neural networks. 
	\end{quotation}

	\section{Introduction}

	A sequential generation of patterns by a system of oscillators is motivated by central pattern generators of biological neural networks \cite{Hooper, brocard}. Central pattern generators usually are considered in the context of neuroscience as an explanation of how nervous systems produce movements. They are autonomous neural networks that can endogenously produce rhythmically patterned output like breathing, walking, or heartbeat. Apart from improving an understanding of the biological aspects, nowadays the engineering aspect of networks of oscillators plays a prominent role  \cite{Manrubia}. For example, complex dynamical structures in populations of phase oscillators were engineered by means of nonlinear time-delayed feedback that is implemented in the interactions of the oscillators \cite{kori}.

	In this work we address the problem of generation of periodic sequences of patterns by a dynamical system of phase oscillators. We consider the case where a set of $P$ patterns $\xi$ should be presented periodically in a certain time order. To solve this problem we design a system which consists of a pacemaker, a retrieval network, and $P$ stored patterns between these elements.
	The proposed system can be seen as an associative memory driven by a pacemaker. Here, we use a particular case of an associative memory (retrieval network) with time-dependent couplings controlled by the pacemaker through the patterns $\xi$. Associative memory models based on the use of phase oscillators are alternatives to the Hopfield model \cite{Hopfield} with spins replaced by phase oscillators. These models are generalizations of the Kuramoto model \cite{kuramoto} (for a review see Acebron \cite{Acebron}). Examples for such generalizations were considered by Aoyagi \cite{Aoyagi} and further improved by Nishikawa {\it et al.} \cite{Nishikawa} with the goal to achieve a storing capacity similar to that of the Hopfield model along with error-free retrieval.

	As a result, our system is able to produce sequences of patterns in the retrieval network. These patterns are encoded as phase differences between the oscillators. The associated dynamical states correspond to phase-locked synchronization. Such states amount to cluster partitions of the oscillators' set, in which oscillators sharing the same phase are gathered in one cluster. The system can be seen as a stylized version of  the central pattern generator of the leech heartbeat \cite{Hooper}, that is composed of two sets, the rhythm generator (corresponding to our pacemaker), and the pattern generator (the retrieval network).

	The paper is organized as follows. In section \ref{sec1} we present the model for our device and its different building blocks. In section \ref{sec2} we illustrate its performance and point on the analogy to a central pattern generator. The conclusions are drawn in section \ref{sec3}.

	\section{The model of our device}\label{sec1}

	The system is composed of a phase oscillator, which plays the role of pacemaker and is characterized by a phase variable $\psi$, a retrieval network with $N$ globally coupled phase oscillators with phases $\phi_i$, $i=1,...,N$, and $P$ patterns $\xi$ stored in the couplings between the pacemaker and the retrieval network.

	The function of the pacemaker is to control the timing and activation of the stored patterns $\xi$ as a function of its phase. In this way, the pacemaker works as a clock that points to the pattern that should be activated in the sequence. The retrieval network encodes the stored patterns $\xi$ as phase differences between its $N$ phase oscillators, one by one as a function of time. We call it retrieval network since its function consists in retrieving the information stored in the couplings between the pacemaker and the network, and translating the patterns to the corresponding cluster partitions of the phase oscillators.

	The operation of the retrieval network is achieved by designing a suitable energy function $L$. The minima of this function are approached by a gradient dynamics in which the energy function plays the role of the potential. The minima correspond to specific cluster partitions, for which the phase differences take only values of zero or $\pi$. These configurations are easily mapped to binary sequences. In the set of $N$ oscillators only $N-1$ phases are independent, therefore we can have $2^{N-1}$ cluster partitions. A pattern $\xi$ is then described as an $N-1$-dimensional vector with components $\xi_i$ being the phase differences between the oscillators $i$ and $N$.

	Now, in order to retrieve a selected pattern out of the $2^{N -1}$ ones that are in principle available, we implement couplings between the pacemaker and the phase oscillators which modulate the energy landscape in a way that the selected pattern becomes the only minimum at a given time. So the proposed system can be seen as an associative memory. The associative memory is particular in the sense that the couplings are time-dependent, driven and controlled by the pacemaker through the choice of patterns $\xi$. For the selection of another pattern, another set of couplings is addressed by the pacemaker. This way it becomes possible to retrieve a whole temporal sequence.

	If the retrieval network had only static sets of couplings between the oscillators, several minima could coexist and would be reached by the gradient dynamics depending on their basins of attraction. This case corresponds to an associative memory network as proposed by Aogagi \cite{Aoyagi} and Nishikawa {\it et al.} \cite{Nishikawa}. Our aim is, however, different. We consider time-dependent couplings in order to have a time-dependent energy landscape that is controlled by the pacemaker. In addition, we impose the constraint that only one minimum of $L$ may exist at a time, when a pattern is selected by the pacemaker. The advantage then is that the gradient dynamics will lead to the required minimum starting from any initial condition due to the absence of competing basins of attraction.

	In the following we shall describe the building blocks of our device in detail, before we illustrate how it  works for retrieving a cyclic sequence of patterns.

	\subsection{The retrieval network}\label{sec1a}

	The dynamics of this network is characterized by an energy function $L$ ("L" shall remind to its role as a Lyapunov function) whose gradient determines the phase evolution of the oscillators. In general, the system evolves to the different minima of $L$ depending on the initial condition and on the basis of attraction of each minimum. For this retrieval network we use a particular version of the model proposed by Nishikawa {\it et al.} \cite{Nishikawa}. We shall show how we tune the couplings in order to have only one minimum in $L$ when a pattern is selected for retrieval. The function $L$ then depends on $N$ oscillator phases $\Phi_i$ in the following way:
	\begin{equation}
	L= -\frac{K}{4N} \sum_{i,j=1, i\not=j}^N \Big( \cos(\Phi_j-\Phi_i) -f_{ij} \Big)^2
	\label{eq1}	
	\end{equation}
	with $\Phi_i \in [0,2\pi[, i=1,...,N$ the phase variables and $K$ the coupling strength that finally determines the speed of convergence of the dynamics towards the stationary state, see Eq. \ref{eq2} below. As noted before, out of the maximally $N(N-1)/2$ different phase differences, only $N-1$ are independent, so we choose $\Delta\Phi_{iN}=\Phi_i-\Phi_N$, $i=1,...,N-1$ as independent variables, using the phase of the $N$th oscillator as reference.  With $f_{ij}$ we denote the couplings, $f_{ij}\in \mathbf{R}, i,j\in \{1,...,N\}$.

	For vanishing couplings $f_{ij}$ we get back the Kuramoto model with twice the usual frequency in the interaction between the oscillators. Then, $L$ has obviously $2^{N-1}$ minima given by the vectors $\xi^{(k)} , k=1,..., 2^{N-1}$ of equal height with components $\xi^{(k)}_i$ $\in\{0,\pi\}$. In the configuration space of phase differences $\{\Delta\Phi_{iN}\}$, these minima are located at the corners of a hypercube of linear size $\pi$ with one corner in the origin of coordinates. It is these $2^{N-1}$ local minima that are our candidates for retrieval. The label $k$ of the state $\xi^{(k)}$ is determined by the pattern of $0$s and $\pi$s interpreted as binary sequence in decimal representation.

	In general, the $f_{ij}$ are couplings appropriately chosen to modulate the energy function $L$ in order to retrieve a selected pattern $\xi^{(s)}$. A sufficient condition for a local minimum reads that the Hessian matrix of $L$ is positive definite, i.e., all eigenvalues being larger than zero, due to an appropriate choice of $f_{ij}$.
	
	Now let us select one pattern $\xi^{(s)}$ with $s\in\{1,...,2^{N-1}\}$. Let the couplings modulate the interaction between oscillator pairs $(ij)$ according to
	\begin{equation}
		f_{ij}(\alpha, s)=\alpha\left(\frac{2}{\pi}\vert\xi_i^{(s)}-\xi_j^{(s)}\vert-1\right)\;
		\label{eq4}
	\end{equation}
	for $i,j\in\{1,...,N,i\not=j\}$ with $\xi_N^{(s)}\equiv 0$, $\alpha$ any real number with $\alpha>1$ and $s$ the index of the selected pattern. So the couplings $f_{ij}$ take values of $\pm\alpha$, depending on the pattern. This choice is guided by the postulate that the selected configuration remains the only local minimum when these couplings are applied, while all other former local minima become saddles in at least one direction or local maxima. The conjecture is that a choice according to Eq.\ref{eq4} satisfies this postulate. In the appendix we shall show that for this choice of couplings the Hessian is positive definite for each configuration that is selected from the $2^{N-1}$ patterns, and not positive definite for all other $2^{N-1}-1$ configurations, which were formerly also minima for vanishing couplings. (For $\alpha<1$ it can be shown that all local minima remain stable, but the selected one becomes the deepest.) The gradient dynamics will then retrieve the local minimum from any initial condition, not necessarily close to the selected minimum.

	It should be noticed that the choice of couplings $f_{ij}$ for a selected pattern $\xi$ by Eq. \ref{eq4} can be mapped to the Hebbian rule in the particular case where only one pattern is memorized by the associative memory of Nishikawa {\it et al.} \cite{Nishikawa}. However, this restriction to the case of only one minimum of the energy function (ensured by the choice of $\alpha>1$) has the advantage that it leads to an error-free retrieval starting from an arbitrary initial condition, possibly far away from the final configuration.
	
	Explicitly the gradient dynamics of the oscillators then reads:
	\begin{equation}
	      \dot{\Phi_i}= -\frac{\partial L}{\partial \Phi_i}= -\frac{K}{N}\sum_{j=1 , j\not=i}^{N} \sin(\Delta\Phi_{ji}) \Big( \cos(\Delta\Phi_{ji})-f_{ji} \Big).
	      \label{eq2}
	\end{equation}
	
	Similarly to the Kuramoto dynamics the interaction of the oscillators depends only on phase differences $\Delta \Phi_{ij}$, and due to the choice of trigonometric functions the interaction terms are bounded as in the Kuramoto dynamics. Due to the gradient dynamics the phase differences will evolve to a fixed point which is the minimum of the energy function $L$ that is closest to the initial conditions. Note that the second order Fourier term was related to the formation of clusters by Mato \cite{Mato}. In our case, this term appears due to the very construction of the energy function.

	\subsection{The role of the pacemaker}

	We extend our dynamics to sequential pattern retrieval via time dependent couplings controlled by a pacemaker. The pacemaker is a phase oscillator with constant frequency $\omega_R$ and phase $\psi$, whose time derivative is given by
	\begin{equation}\label{equ_clock}
		\dot{\psi} = \omega_{R}.
	\end{equation}

	It is the phase $\psi$ that selects and activates a pattern $r$ among the $P$ stored ones over a duration $B$. This activation period lies in the phase (time) interval between $\psi=\psi_r$ and $\psi=\psi_{r+1}$. For simplicity we have chosen $\psi_r=\frac{2\pi}{P}(r-1)$ for all $r\in\{1,...,P\}$, and $B =\frac{2\pi}{P}$.
	This means, when the instantaneous phase $\psi$ comes to the value $\psi_r$, couplings $f_{ij}$ are switched on that guarantee the retrieval of the pattern $^r\xi^{s(r)}$. Here the pre-superscript $r$ indicates the label of the pattern within the time sequence, the post-superscript $s(r)$ stands for the decimal label of the pattern $r$ out of the selected subset. The corresponding dynamical equations read:
	\begin{equation}
		\dot{\Phi_i}\;=\; -\frac{K}{N}\sum_{j=1, j\not=i}^{N} \sin(\Delta\Phi_{ji}) \Big( \cos(\Delta\Phi_{ji})-f_{ji}(\psi) \Big),
		\label{equ_oscillators}
	\end{equation}
	where
	\begin{equation}
		f_{ij}(\psi) = \sum^{P}_{r=1} \alpha\left(\frac{2}{\pi}\vert ^r\xi_i^{s(r)}- \; ^r\xi_j^{s(r)}\vert\;-\;1\right)g_{r}(\psi)
		\label{couplings_psi}
	\end{equation}
	and
	\begin{equation}
		g_r(\psi) = \Theta(\psi - \psi_r) - \Theta(\psi - \psi_r - B).
		\label{g}
	\end{equation}

	Here $\Theta$ denotes the Heavyside function. This means that the function $g_r(\psi)$ controls the couplings to be given as $\alpha\left(\frac{2}{\pi}\vert ^r\xi_i^{s(r)}-\; ^r\xi_j^{s(r)}\vert-1\right)$ over the phase interval $B$, starting from $\psi = \psi_r$ on.

	Obviously we should ensure that $\psi_{r+1}-\psi_r\ge B\ge \psi_{trans}$, that is, the time interval between two initiations of pattern retrievals and the time of application of the constant couplings (to reach the new pattern) should be larger than the transient time $\psi_{trans}$ which the retrieval network needs to go from one pattern to the next (the speed of the internal dynamics is controlled by $K$).
	
	The information for the sequence generation is stored in the phase values $\psi_r$'s through the functions $g_r(\psi)$ which ``switch on" the appropriate couplings. Since the stored patterns $\xi$ are required at different times, they must be stored outside the retrieval network. Therefore we say that the patterns generated in the sequence are memorized in the couplings between the pacemaker and the retrieval network. This way the architecture can modulate the couplings $f_{ij}$ between the oscillators of the retrieval network.

In summary, the pacemaker transforms the static contents, memorized in the couplings, into a temporal sequence of patterns, a feature that is in common with a central pattern generator.

	To complete our set of equations, we have to introduce noise in the retrieval dynamics. The reason is the following. As we have seen, once a pattern is selected for retrieval, the system evolves according to the dynamics of Eq. \ref{eq2} to the only stable fixed point of $L$ of the retrieval network. When the next pattern of the sequence is activated by the pacemaker, the system is still in the former fixed point that turns into an unstable one. To kick the system out of this fixed point and follow the required sequence, we apply Gaussian white noise of small intensity T. Eq.\ref{eq2} is then replaced by

	\begin{equation}
		\dot{\Phi_i}\;=\; -\frac{K}{N}\sum_{j=1, j\not=i}^{N} \sin(\Delta\Phi_{ji}) \Big( \cos(\Delta\Phi_{ji})-f_{ji}(\psi) \Big) + T\eta_i(t),
		\label{equ_oscillators_noise}
	\end{equation}
	where $\eta_i(t)$ is a random variable describing the white noise with zero mean, ͗$<\eta_i(t)\eta_j(t')> = \delta_{ij} \delta(t-t')$, and $T$ is the noise intensity. Our system is then described by Eq.s \ref{equ_clock}, \ref{equ_oscillators_noise}, \ref{couplings_psi}, and \ref{g}.

	\section{Numerical study}\label{sec2}

	In this section we study an example of this system and focus on its dynamical properties.
	  We integrate the dynamics using a second order stochastic Runge-Kutta method\cite{Honeycutt}  with time step $\Delta t = 0.01$ and noise intensity $T=0.001$. We consider a set of $N=11$ phase oscillators in the retrieval network. This network can therefore encode $2^{10} = 1024$ patterns as cluster partitions. Out of this set, we have randomly selected $P=5$ patterns. They are $ \;^1\xi^{672}$, $\; ^2\xi^{0}$,$\; ^3\xi^{942}$, $\; ^4\xi^{477}$, $\; ^5\xi^{1023}$ in the indicated order. The time dependent-couplings are changed at phase values $\psi_r=0,\; 2\pi/5,\; 4\pi/5,\;6\pi/5$ and $8\pi/5$. The parameters are chosen as $K = 10$, $B=2\pi/P$, $\alpha =2$ and $\omega_R=1$.

\begin{figure}[!ht]
	\begin{center}
	\includegraphics[width=1.0\columnwidth]{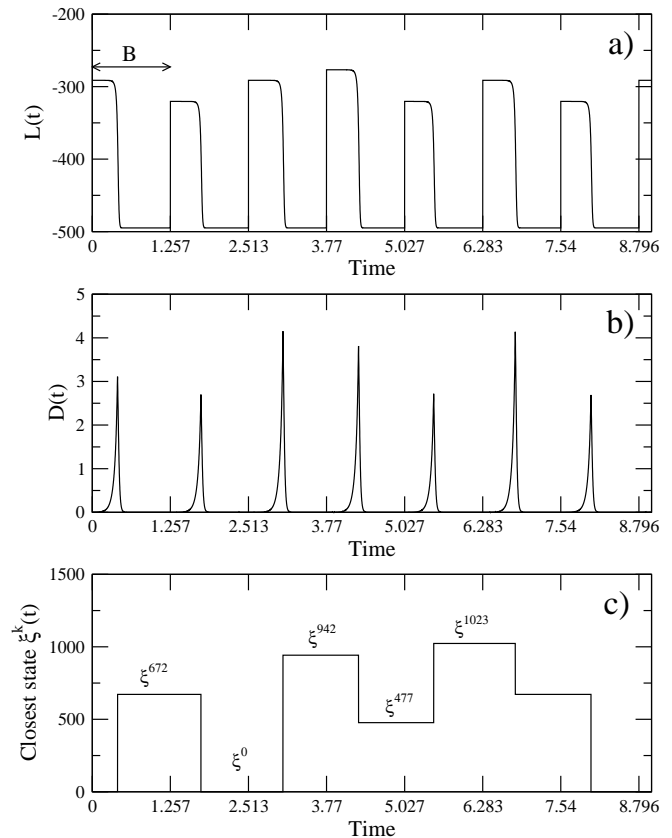}
		\end{center}
		\caption{Sequential pattern retrieval of a sequence $s_1$ =  $\{ \xi^{672}$, $\xi^{0}$, $\xi^{942}$, $\xi^{477}$, $\xi^{1023} \}$. (a) Energy $L(t)$ as function of time. (b) Euclidean distance of the actual state of the system to the closest state (corresponding to patterns on the corners of the hypercube). It vanishes for roughly half of the period $B$, so that the system then has retrieved the desired state. (c) Closest states to the current evolving state as function of time. As seen from the figure, the closest states themselves vary with time. The set of closest states agrees with the set of selected states. The states carry their decimal labels.}
		\label{fig3}
	\end{figure}

	Figure \ref{fig3}a shows the energy $L$ as function of time evaluated in the actual state of the retrieval network. $L$ jumps from its minimal value at $L\approx -500$ to some larger value around $L\approx-300$, where it remains as long as the system of oscillators searches the new minimum, corresponding to the new choice of external fields. When the new minimum is found, $L$ drops to the minimal value again. During such an interval of duration $B$, the Euclidean distance $D(t)$ in configuration space between the actual state and the closest pattern (corner of the hypercube) has a peak at an intermediate time interval where the system is moving from one to the next selected state. We see these peaks in Fig.~\ref{fig3}b. For about half of the period $B$ this distance is zero, indicating that the state of the system system corresponds to the required pattern.

	In Fig.~\ref{fig3}c we plot the states which are closest to the instantaneous states of the system as a function  of time. Obviously the closest states are just the selected ones, but this does not mean that the actual states (evolving with time) are identical with the selected ones  over the whole duration of the plateau; as mentioned before,  the distance to the selected states vanishes only for roughly half of the period as it is seen from Fig.~\ref{fig3}b. The width of the peaks in the distance from the closest states can be tuned by the coupling parameter $K$, large $K$ accelerates the convergence to a new pattern, once the time-dependent couplings are changed; also the strength of the couplings, parameterized by $\alpha$, determines the speed of convergence.

	\begin{figure}[!ht]
	      \begin{center}
	      \includegraphics[width=1.0\columnwidth]{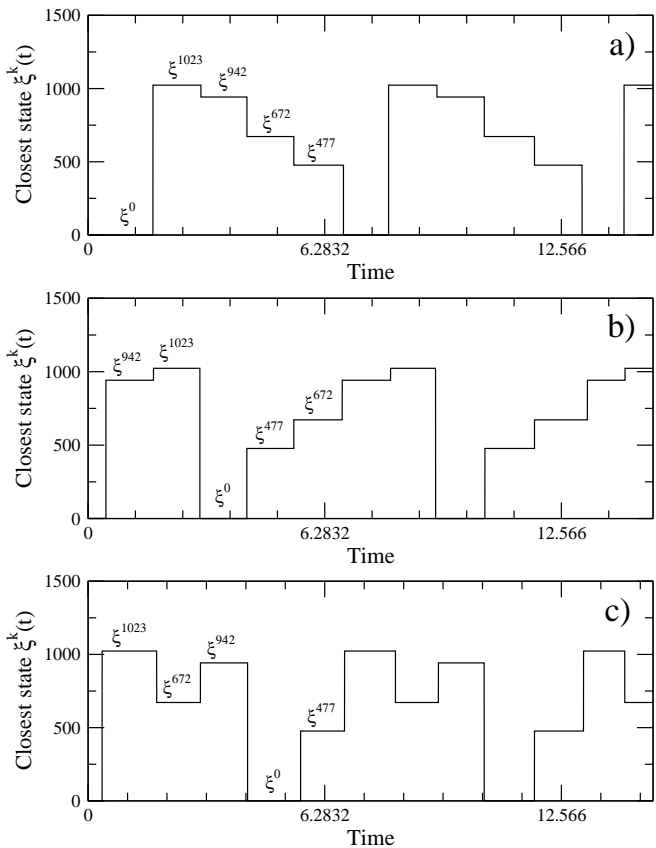}
	      \end{center}
	      \caption{Sequential pattern retrieval of different sequences with the same states. (a) Time evolution of the sequence $s_2$ =  $\{ \xi^{0}$, $\xi^{1023}$, $\xi^{942}$, $\xi^{672}$, $\xi^{477} \}$. (b) Time evolution of the sequence $s_3$ =  $\{ \xi^{942}$, $\xi^{1023}$, $\xi^{0}$, $\xi^{477}$, $\xi^{672} \}$. (c) Time evolution of the sequence $s_4$ =  $\{ \xi^{1023}$, $\xi^{672}$, $\xi^{942}$, $\xi^{0}$, $\xi^{477} \}$. In all cases we show only the states which are closest to the current evolving state. For roughly half a period the distance between the closest 	state and the actual state of the system vanishes, which is interpreted as pattern retrieval.}
		\label{fig4}
	\end{figure}

	In Fig.~\ref{fig4} we show three permutations of the same set of the five stored patterns. Indicated are the five plateaus in time where a certain pattern remains the closest to the current state and where this pattern agrees with the system's state over roughly half of the period (the analogous figures to Fig.~\ref{fig3}b are not displayed here). It should be noticed that a change in the pattern sequence from Fig.~\ref{fig4}a to \ref{fig4}b and \ref{fig4}c only amounts to reorder the phase shifts $\psi_r$, no other change of the system's structure is needed.

	\section{Conclusions}\label{sec3}

	We have designed a system of phase oscillators that is able to produce periodic sequences of patterns. Patterns are stored in the couplings of the system and retrieved and encoded as phase differences. Due to the task division between the pacemaker, the stored patterns and the retrieval network, the system is very flexible and robust. Different sequences of the stored patterns can be implemented without modifying the system's structure. The retrieval network itself operates in a robust way since it has by construction only one minimum, therefore the dynamics converges to the desired pattern independently of the initial conditions.

	Our device may be regarded as a very stylized version of the central pattern generator of the leeches' heartbeat. According to Hooper \cite{Hooper}, the central pattern generator of the leech heartbeat can be divided into two sets, the rhythm generator (corresponding to our pacemaker), and the pattern generator (corresponding to our retrieval network). The pattern generator there generates the actual motor pattern in response to the driving input from the rhythm generator (in our case in response to the driving input of time-dependent couplings from the pacemaker).

	\section{Acknowledgments}
	We would like to thank Alexander S. Mikhailov very much for valuable discussions.

	\section{Appendix}

	Consider the energy function for an all-to-all coupled system of $N$ phase oscillators:

	\begin{equation}\label{aeq1}
		L=-\frac{K}{4N}\sum_{ij, i\not=j}^N (\cos(\Phi_j-\Phi_i) -f_{ij})^2.
	\end{equation}

	From now on we set the coupling strength $K=1$. Let the couplings be chosen according to
	\begin{equation}\label{aeq4}
		f_{ij}(\alpha, s)=\alpha \bigg(\frac{2}{\pi}\vert\xi_i^s-\xi_j^s\vert-1\bigg)\; \textrm{for} \;i,j \in \{ 1,...,N, i \not= j \}
	\end{equation}
	with $\Delta\Phi_{ij}=x_i-x_j$, $\xi_N^s=x_N\equiv 0$, $\alpha$ any real number with $\alpha>1$ and $s$ the index of the selected pattern. We now prove a sufficient condition that the Hessian matrix with respect to the $N-1$ independent phase differences is positive definite for the selected pattern and not positive definite for all other $2^{N-1}-1$ patterns, provided we choose the external fields $f_{ij}$ according to Eq.\ref{eq4}. From the first derivative $\partial L/\partial x_{i}$ we immediately see that candidates for extrema are $x_{i}\in \{0,\pi\}$, where $x_i$ was defined as $\Delta\Phi_{iN}=\Phi_i-\Phi_N$, while for $f_{iN}>1$ the individual cos-dependent terms are different from zero. For a particular given choice of $f_{ij}, f_{iN}$ with possibly alternating signs the first derivatives can vanish also at intermediate values of $x_{i}$ which we project on $[0,2\pi[$ that can lead to further extrema. This part we treat numerically in order to exclude that these extrema compete with the selected minimum that shall be retrieved.

	Next let us consider the Hessian of $L$ as function of the phase differences.  Apart from the normalization factor, its diagonal elements are given as:
\begin{eqnarray}
	\frac{\partial^2 L}{\partial x_{k}^2} &=& -\sin^2(x_{k}) + \cos(x_{k})\big( \cos(x_{k}) - f_{kN} \big)
	  \nonumber \\
	 &-& \sum_{j=1, j \ne k}^{N-1} \bigg( \sin^2(x_{k} - x_{j}) - \cos(x_{k} - x_{j})\\ \nonumber
&\cdot& \big( \cos(x_{k}
        - x_{j}) - f_{jk} \big) \bigg)\;,
	\label{equ_full_deri_22}
	\end{eqnarray}
its off-diagonal elements are:
	\begin{equation}
	\frac{\partial^2 L}{\partial x_{k} \partial x_{l}} =   \sin^2(x_{k} - x_{l}) - \cos(x_{k} - x_{l})\big( \cos(x_{k} - x_{l}) - f_{lk} \big) .
	\label{equ_full_deri_21}
	\end{equation}
For a choice of the external fields $f_{ij}$ according to the rule (\ref{eq4}), the matrix simplifies to
\begin{equation}
	\frac{\partial^2 L}{\partial x_{k}^2} = -\Bigg\{ \mp 1 \big( \pm 1 - f_{kN} \big) + \sum_{j=1, j\not=k}^{N-1} \bigg( \mp 1 \big( \pm 1 - f_{jk} \big) \bigg)\Bigg\}
	\label{equ_full_deri_22_corner}
	\end{equation}
for the diagonal elements and $k=1,...,N-1$, and to
	\begin{equation}
	\frac{\partial^2 L}{\partial x_{k} \partial x_{l}} =  \mp 1 \big( \pm 1 - f_{lk} \big)
	\label{equ_full_deri_21_corner}
	\end{equation}
for the off-diagonal elements, with $k,l\in\{1,...,N-1\}, k\not=l$. The upper (lower) sign in front of the the bracket with $f_{lk}$ stands for the case that the difference of components $\vert x_{k}^s-x_{l}^s\vert$, read off from the selected configuration $\vec{\xi}^s$, is zero ($\pi$), respectively.

\noindent Next we study the positive definiteness of this matrix for all $2^{N-1}$ configurations which may be selected for retrieval. Here it is convenient to classify the configurations in terms of their Hamming distance from the selected pattern, i.e. the number of mismatches of components
between $\vec{\xi}$ and $\vec{\xi}^{s}$, which varies between zero and $N-1$. By a suitable permutation of the coordinate axis in configuration space
we can always achieve that the k mismatches occur in the first k coordinates of $\xi$ so that the corresponding Hessian $H$ is chosen as representative for all patterns with k mismatches.\\\\
 \noindent
 {\bf H with no mismatches}
 According to our choice of $f_{ij}$, their signs are opposite to those of $\cos(\Delta\Phi_{lk}^s)$, that is $\Delta\Phi_{lk}^s=0$ or $(\pi)$, so that $\cos\Delta\Phi_{lk}^s=1$ or $(-1)$ and $f_{lk}=-\alpha$ (or $+\alpha$), $\alpha>1$, respectively.  The diagonal elements then simplify to $\partial^2L/\partial x_{k}^2=(N-1)(1+\alpha)$, the off-diagonal elements to $\partial^2L/\partial x_{k}\partial x_{l}=-(1+\alpha)$. The Hessian therefore takes the form of an $(N-1)\times(N-1)$ dimensional circulant matrix, whose eigenvalues turn out to be $\lambda_1=1+\alpha$ with multiplicity 1 and $\lambda_2=N(1+\alpha)$ with multiplicity (N-2). (Here we have used the following: Eigenvalues of an $n\times n$ circulant matrix, specified by the vector $(c_0,c_1,...,c_{n-1})$, are known to be given as $c_j^\prime=\sum_{k=0}^{n-1}e^{2\pi ijk/n}c_k$ with $j=0,-1,-2,...,-(n-1)$. In our case the Hessian has a particularly simple form, for which one element in each row is $(N-1)(1+\alpha)$, while all other $N-2$ elements are $-(1+\alpha)$. Using these values and the fact that the sum over all n roots of the unit circle adds up to zero leads to our results for the eigenvalues.) Now, since for $\alpha>1$ all eigenvalues are positive, the selected configuration corresponds to a local minimum in configuration space, whatever pattern has been chosen for retrieval. (In order to have only a local minimum at the selected configuration, obviously $\alpha>-1$ would be sufficient, but at the same time, the other patterns should become saddles or local maxima, and in view of that we shall need $\alpha>1$, see below.)\\\\
 \noindent {\bf H with one mismatch}
 Next we evaluate the Hessian for a configuration that differs from the selected pattern in a single phase difference. Without loss of generality we assume the mismatch to happen in the first coordinate, affecting the Hessian in the first column and the first row according to $H_{11}=(N-1)(1-\alpha)$, $H_{1j}=(\alpha-1)=H_{j1}$ for $j=2,...,N-1$, while the remaining $(N-2)(N-2)$ submatrix remains circulant. The Sylvester criterion, applied to the positive definiteness of the overall $(N-1)\times(N-1)$ matrix,  is now violated due to the first element $H_{11}=(N-1)(1-\alpha)<0$ for $\alpha>1$, so that the configuration with one mismatch is no longer a local minimum of the energy function $L$. (As necessary and sufficient condition for a Hermitian matrix  to be positive definite, the Sylvester criterion requires that all leading principal minors of the matrix are positive.)\\\\
 \noindent{\bf H with $k>1$ mismatches}
 Now the configuration has $k$ mismatches with the selected configuration which we arrange to occur in the first $k$ coordinates. Here it should be noticed that $f_{iN}$ will have the ``wrong" sign with respect to $\Delta\Phi_{iN}$, $i=1,..,k$, but $f_{il}$ will have the ``right" sign with respect to $\Delta\Phi_{il}$ for $i,l\in\{1,...,k\}$, since two mismatches compensate in the relative phase differences (``wrong" (or ``right") refer to the feature which prevents (or ensures) the property of becoming a local minimum, respectively.) This explains why the components of the $k\times k$ submatrix $S_k(H)$ in the upper left corner of the Hessian are given by
 \begin{equation}
 S_k(ii)=(N-1) - \alpha(N-2k+1),\;\; i=1,...,k
 \end{equation}
 for the diagonal elements and
 \begin{equation}
 S_k(ij)= -(1+\alpha)\;\;  i,j=1,...,k,\;,i\not=j
 \end{equation}
 for the off-diagonal elements.
The submatrix $S_k(H)$ is again circulant and has eigenvalues $\lambda_1=(N-k)(1-\alpha)$ with multiplicity 1 and $\lambda_2=N-\alpha(N-2k)$ with multiplicity $k-1$, so that the determinant of this submatrix reads $\vert S_k(H)\vert=\lambda_1\lambda_2^{k-1}$. Now we have to distinguish the following cases:\\
 \noindent 1. k odd. For k odd, $\lambda_2^{k-1}$ is always positive while $\lambda_1<0$ for $\alpha>1$, so that $\vert S_k(H)\vert<0$ for odd k and $\alpha>1$ and the Sylvester criterion for $H$ being positive definite is violated as it should be for any positive number of mismatches.\\
 \noindent 2. k even. For k even, both eigenvalues may be negative so that $\vert S_k(H)\vert>0$. In order to see that the Sylvester criterion is still violated, we have to distinguish the following cases: \\
 \noindent (i) For $\alpha>1$ and $k>N/2$ we have $\lambda_1<0$ and $\lambda_2>0$, so that the Sylvester criterion is violated. \\
 \noindent(ii) For $\alpha>1$ and $k<N/2$, $\lambda_2<0$ for $\alpha>N/(N-2k)$, so that $\vert S_k(H)\vert<0$ only for $1<\alpha<N/(N-2k)$.\\
 \noindent
 (iii) To finally see what happens for $\alpha>1$ and $\alpha>N/(N-2k)$ let us consider the determinant of the submatrix of size $l=k-1$ in the upper left corner of H. This matrix has eigenvalues $\sigma_1=(N-k+1)-\alpha(N-k-1)$ and $\sigma_2=N-\alpha(N-2k)$ with even algebraic multiplicity $(k-2)$, so that again the sign of $\lambda_1$ determines the sign of this subdeterminant. Now $\sigma_1<0$ for $1<\frac{N-k+1}{N-k-1}<\alpha$, but this is certainly satisfied, since in the considered case $k\geq 2$ and $\alpha$ was even larger than $N/(N-2k)$ by assumption. So this $(k-1)\times(k-1)$-dimensional subdeterminant violates the Sylvester criterion for H to be positive definite.

 In particular, for the maximal number of mismatches $k=N-1$, $\lambda_1=1-\alpha<0$ for $\alpha>1$ and $\lambda_2=N+\alpha(N-2)>0$ for $N>2$, and for $N=2$, $\lambda_2^{k-1}=\lambda_2^{N-2}=1>0$, so that the corresponding pattern again ceases to be a local minimum of the energy function.


\end{document}